\documentclass[pra,twocolumn,aps,preprintnumbers,amsmath,amssymb,superscriptaddress]{revtex4-2}

\usepackage{color}
\usepackage{graphicx}
\usepackage{epstopdf}
\usepackage{epsfig}
\usepackage{dcolumn}
\usepackage{bm}
\usepackage{hyperref}
\usepackage{mathrsfs}
\usepackage{comment}
\usepackage{soul}
\usepackage[normalem]{ulem} 
\usepackage{ulem}
\usepackage{xcolor}
\usepackage{amsfonts}
\usepackage{amsmath}
\usepackage{amsxtra}
\usepackage{amssymb}
\usepackage{mathrsfs}
\usepackage{hyperref}
\usepackage{soul}
\usepackage{float}
\usepackage{relsize}
\usepackage{placeins}
\usepackage[T1]{fontenc}
\begin{document}


\title{Photon loss effects on light-mediated non-Gaussian entangled Bose-Einstein condensates projecting with different photon measurement outcomes}

\author{Shuai Gao}
\affiliation{Joint International Research Laboratory of Information Display and Visualization, School of Electronic Science and Engineering, Southeast University, Nanjing 210096, China}

\author{Manish Chaudhary}
\affiliation{State Key Laboratory of Precision Spectroscopy, School of Physical and Material Sciences, East China Normal University, Shanghai 200062, China}
\affiliation{New York University Shanghai, 567 West Yangsi Road, Shanghai, 200126, China}

\author{Alexey N. Pyrkov}
\affiliation{Federal Research Center of Problems of Chemical Physics and Medicinal Chemistry of Russian Academy of Sciences, Acad. Semenov av. 1, Chernogolovka, Moscow Region 142432, Russia}

\author{Ebubechukwu O. Ilo-Okeke} 
\affiliation{New York University Shanghai, 567 West Yangsi Road, Shanghai, 200126, China} 
\affiliation{Department of Physics, School of Science, Federal University of Technology, P. M. B. 1526, Owerri, Imo state 460001, Nigeria}

\author{Xin Meng}
\affiliation{State Key Laboratory on Tunable Laser Technology, Harbin Institute of Technology (Shenzhen), Shenzhen 518055, China}
\affiliation{Guangdong Provincial Key Laboratory of Aerospace Communication and Networking Technology, Harbin Institute of Technology (Shenzhen), Shenzhen 518055, China }

\author{Jingyan Feng}
\affiliation{State Key Laboratory of Precision Spectroscopy, School of Physical and Material Sciences, East China Normal University, Shanghai 200062, China}

\author{Muhammad Jamil Khan  }
\affiliation{Key Laboratory of MEMS of Ministry of Education, Southeast University, Nanjing 210096, China}

\author{Tim Byrnes}
\affiliation{State Key Laboratory of Precision Spectroscopy, School of Physical and Material Sciences, East China Normal University, Shanghai 200062, China}
\affiliation{NYU-ECNU Institute of Physics at NYU Shanghai, 3663 Zhongshan Road North, Shanghai 200062, China; Shanghai Frontiers Science Center of Artificial Intelligence and Deep Learning, NYU Shanghai, 567 West Yangsi Road, Shanghai, 200126, China.}
 \affiliation{Center for Quantum and Topological Systems (CQTS), NYUAD Research Institute, New York University Abu Dhabi, UAE.}
\affiliation{Department of Physics, New York University, New York, NY 10003, USA}

\author{Chaogang Lou}
\email{lcg@seu.edu.cn}
\affiliation{Joint International Research Laboratory of Information Display and Visualization, School of Electronic Science and Engineering, Southeast University, Nanjing 210096, China}

\date{\today}

\begin{abstract}

The theory of quantum information processing for macroscopic qubits is based on the fact that every macroscopic qubit has a conserved number of particles. However, from an experimental point of view, every such qubit experiences processes of decoherence that impact the possibilities for entanglement generation between such qubits and use in quantum information processing efficiently. One of the most prospective methods for generating entanglement between distant atomic BECs is quantum nondemolition measurements. Here, we study how the effects of photon measurement impact the entanglement when photon loss decoherence is included. We employ the thermally entangled state representation (TESR) and integral within the ordered operator(IWOP) approach to obtain the accurate density matrix in a photon loss channel. We demonstrate that varying outcomes of photon number measurements lead to the generation of distinct entangled states, each exhibiting unique characteristics. We find that using the Hofmann-Takeuchi and Duan-Giedke-Cirac-Zoller criterion provides advantages in entanglement detection compared to the Wineland squeezing and EPR steering criterion in such settings. 
\end{abstract}

 \maketitle

\section{Introduction}
Quantum squeezing is one of the main techniques in the advancement of quantum metrology and quantum information science \cite{Walls1983Squeezed,scully1999quantum,nielsenchuang,gerry_knight_2004,PhysRevLett.131.110602}. According to the Heisenberg uncertainty principle, it is conceivable to reduce the quantum mechanical noise (measured by the variance) of one operator while simultaneously increasing it for another operator in a squeezed state \cite{PhysRevD.4.1925,PhysRevA.47.5138,2014Generation}. Consequently, spin squeezed states have been employed to enhance the sensitivity beyond the conventional quantum limit \cite{2008Squeezing}. These states offer a wide range of practical applications in optical interferometry \cite{Dowling,2017SqueezedSchnabel}, magnetometry and electric-field sensing \cite{PhysRevX.5.031010,Fan_2015}, and gravitational wave detection \cite{Oelker:14,PhysRevLett251102}. The first demonstrations of two-mode squeezed states for atomic
ensembles were pioneered by the group of Polzik and co-workers using the Holstein-Primakoff approximation $x=S^{y} / \sqrt{2 N} \text { and } p=S^{z} / \sqrt{2 N}$ \cite{krauter2011entanglement,RevModPhys.82.1041,julsgaard2001experimental,kong_measurement-induced_2020,PhysRevLett.92.030407,PhysRevLett.85.5643,PhysRevLett.85.5639}. In this method, the discrete spin operators are converted into the continuous position and momentum operators of a harmonic oscillator.
Additionally, these spin squeezing states serve as a valuable tool \cite{2008Squeezing,julsgaard2001experimental,PhysRevLett.92.030407} for facilitating the detection of entanglement in complex spin ensembles \cite{PhysRevA.104.053324,PhysRevA.108.032420}.


Quantum nondemolition (QND) measurements have become a well-established technique in the field of quantum mechanics to measure and manipulate quantum mechanical systems. This entangling scheme has been demonstrated as a highly efficient method for generating squeezing in various systems, as evidenced by a range of studies \cite{PhysRevX.5.041037,kuzmich2000generation,higbie2005,doi:10.1126/science.209.4456.547,PhysRevA.60.4974,PhysRevA.104.053324,PhysRevLett.100.103601,2009Implementation,sewell2012magnetic,doi:10.1126/science.aaf3397,PhysRevLett.116.093602}. QND measurements are regarded as a form of measurement that minimally perturbs the system considering the interaction between atoms and light is relatively weak, resulting in a restricted collapse of the wave function of the atoms. For Bose-Einstein condensates (BECs), the creation of many-particle entanglement localized in a single spatial location \cite{Gross_2012,2001Many,Augusto2014Fisher,PhysRevLett.86.4431} and spatially separate regions \cite{kunkel2018spatially,fadel2018spatial,lange2018entanglement,2013Entanglement} within one BEC condensate has been observed. The entanglement between two spatially distinct BECs was realized in an experiment recently \cite{PhysRevX.13.021031}.

In a previous study, we investigated the effects of atom dephasing on the entanglement between Bose-Einstein condensates (BECs) induced by QND \cite{AristizabalZuluaga2021QuantumNM,Gao_2022}. Based on QND-induced entanglement under various sources of decoherence, a protocol for generating entanglement between BECs was proposed. The present paper focuses primarily on the effects of photon measurement under a photon loss channel, which influences the number of photons that can be entangled with BECs. The experimental protocol we employ differs from the previous approach \cite{AristizabalZuluaga2021QuantumNM,Gao_2022}, wherein two BECs are positioned on the arms of the Mach-Zehnder interferometer. In contrast, our methodology involves the continuous passage of light through the BEC to acquire distinct Hamiltonians \cite{pettersson2017light}.

.

The structure of this paper is as follows. Sec. \ref{ii} provides a concise overview of the QND entangling protocol, elucidating the underlying physical framework and outlining the fundamental system under investigation. In Section \ref{photon_loss_chap}, we analyze the density matrix master equation and study the evolution of the entangled state under the photon loss channel, evaluating its entanglement using different criteria.  Finally, the results are summarized and discussed in Sec. \ref{conclusion}.

\begin{figure}[t]
\includegraphics[width=\linewidth]{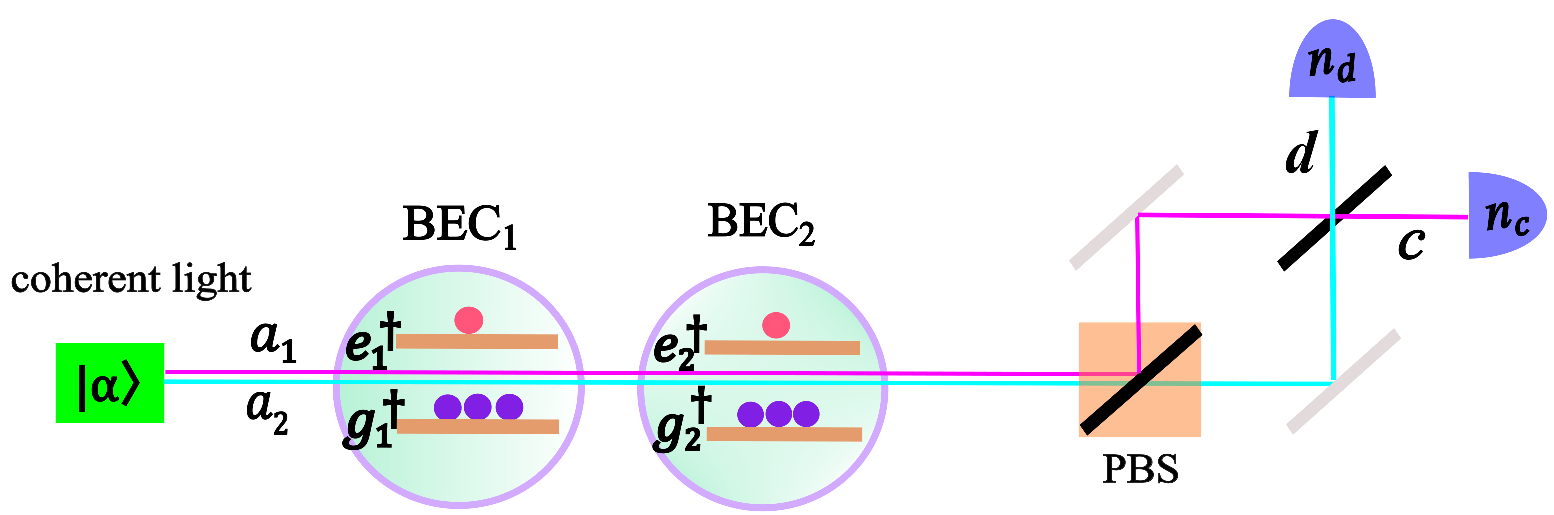}
\caption{\label{Experimental_scheme}Experimental scheme for generating entanglement with QND measurement. A coherent light pulse $|{\alpha}\rangle$ entangles BECs and light while passing through two BECs initially in two spin coherent states prepared in the $S^x$ direction. The light modes $a_1,a_2$ are then converted to another two light modes by a polarizing beam splitter (PBS).  After another two beam splitters and photon number measurement, the atomic state evolves to an entangled state.}
\end{figure}

\section{QND INDUCED ENTANGLEMENT PROTOCOL} \label{ii}

\subsection{Physical realization scheme and Hamiltonian}
\label{ii.A}

We first briefly introduce the QND entangling scheme shown in Fig.\ref{Experimental_scheme}. One reasonable choice for Bose-Einstein condensates (BECs) could be separated magnetic traps on an atom chip or two optical dipole traps \cite{reichel2011atom,whitlock2009t,abdelrahman2014coherent}. Each BEC has two internal energy states and the corresponding bosonic annihilation operators are given by $g_j,e_j$ where $j \in \{1,2 \}$  that labels different BECs. All the internal energy states are populated by atoms and a suitable candidate for $^{87} \text{Rb}$ could be two hyperfine ground states where $F = 1/2$. The system is initially illuminated by coherent light, and the initial state of optical mode $a1,a2$ can be described as

\textbf{\begin{equation}
\label{light coherent state}
|\alpha \rangle \equiv e^{ -|\alpha|^2 /2 } e^{\big[\frac{\alpha}{\sqrt{2}} (a_1^\dagger+a_2^\dagger) \big]} | \mathrm{vac} \rangle. 
\end{equation}}

To simplify the calculation, we use Schwinger boson operators to define effective spin for atoms, 
\begin{align}
\label{schwinger}
S^x_j & = e^{\dag}_j g_j + g^{\dag}_j e_j, \nonumber \\
S^y_j & = -i e^{\dag}_j g_j + i g^{\dag}_j e_j, \nonumber \\
S^z_j & = e^{\dag}_j e_j - g^{\dag}_j g_j . 
\end{align}
which follows the relative communication relations  $ [S^j, S^k] = 2i \epsilon_{jkl} S^l $, and $ \epsilon_{jkl} $ is the completely anti-symmetric Levi-Civita tensor.

We prepare the initial state of the atoms in the spin coherent states, which are polarized in a particular direction defined by angles $ 0 \leq \theta \leq \pi$, $ -\pi \leq \phi \leq \pi$ on a Bloch sphere.

\begin{align}
|\theta, \phi  \rangle \rangle_j &\equiv \frac{1}{\sqrt{N!}} \left( e^\dag_j \cos \frac{\theta}{2} 
 + e^{i\phi} g^\dag_j \sin \frac{\theta}{2} \right)^N | \text{vac} \rangle \nonumber \\
& = \sum_{k=0}^{N}\sqrt{\binom{N}{k}}\cos^{k}\left(\frac{\theta}{2}\right)\sin^{N-k}\left(\frac{\theta}{2}\right)e^{i(N-k)\phi } |k\rangle_j .
\label{coherent state expression}
\end{align}
and the Fock states are defined by
\begin{align}
|k\rangle_{j}=\frac{\left(e_{j}^{\dagger}\right)^{k}\left(g_{j}^{\dagger}\right)^{N-k}}{\sqrt{k !(N-k) !}}|\mathrm{vac}\rangle
\label{eq: fockstate}                                                       \end{align}

where $|\mathrm{vac}\rangle$ is the vacuum state which has no photons or atoms. The form of the Hamiltonian that interacts light with atoms of two BECs is as follows:
\begin{align}
\label{Hamilton new}
H=q\left(S_{1}^{z}+S_{2}^{z}\right) 
L^{z} .
\end{align}
where we use the Stokes operator $L^z=a_{1}^{\dagger} a_{1}-a_{2}^{\dagger} a_{2}$ to represent optical modes, and $q=\frac{\sigma \gamma \alpha_{v}}{A\left(I+\frac{1}{2}\right) \Delta}$ is the interaction strength parameter originated from ac Stark shift coupling. In this context, $\sigma$ represents the resonance absorption cross-section for the interaction between an unpolarized photon and an unpolarized atom, $\gamma$ denotes the rate of spontaneous emission from the upper atomic level, and $A$ depicts the cross-sectional area of the light beam. Additionally, $\alpha_{v}$ signifies the vector polarizability and $I$ represents the nuclear spin value \cite{PhysRevLett.85.5639,pettersson2017light}.

\subsection{ QND wavefunction}
\label{entanglement wavefunction}

The initial state of the BECs is prepared in the $S^x$ polarized direction on the Bloch sphere
\begin{align}
|\tilde{\Psi}_0 \rangle &= \left. \Big|\frac{\pi}{2},0 \Big\rangle \Big\rangle_1 \right.\left. \Big|\frac{\pi}{2},0 \Big\rangle \Big\rangle_2\right. \nonumber \\ & = |k_x=N \rangle_1|k_x=N \rangle_2
\nonumber \\ & = \sum_{k_{1}, k_{2}=0}^{N} \frac{1}{2^{N}} \sqrt{\binom{N}{k_1} \binom{N}{k_2} 
}\left|k_{1}\right\rangle\left|k_{2}\right\rangle 
\label{becinitial}
\end{align}

The light passes through the two BECs successively, producing an interaction with a duration of time $t$, and the system is resolved using the time evolution operator $\text{exp}^{-itH/\hbar}$,

\begin{align}
|\tilde{\Psi}(\tau)\rangle &=  \exp (-i t H / \hbar)|\tilde{\Psi}_0 \rangle\left|\alpha\right\rangle  \nonumber \\
= & \frac{1}{\sqrt{2^{\left(N_{1}+N_{2}\right)}}} \sum_{k_{1}, k_{2}} \sqrt{C_{N_{1}}^{k_{1}} C_{N_{2}}^{k_{2}}} \nonumber \\
& \times e^{-i \tau\left[2 k_{1}+2 k_{2}-\left(N_{1}+N_{2}\right)\right]\left(a_1^{\dagger} a_1-a_2^{\dagger} a_2\right)}\left|k_{1}, k_{2}\right\rangle|\alpha\rangle
\label{initial density}
\end{align}

Where the dimensionless time $\tau=qt/\hbar$. Noting that $e^{i \theta c^{\dagger} c} e^{\alpha c^{\dagger}}|0\rangle=\exp \left[\alpha e^{i \theta} c^{\dagger}\right]|0\rangle$, we may rewrite the above expression in the following form

\begin{align}
|\tilde{\Psi}(\tau)\rangle= & \frac{e^{-\frac{|\alpha|^{2}}{2}}}{\sqrt{2^{\left(N_{1}+N_{2}\right)}}} \sum_{k_{1}, k_{2}} \sqrt{C_{N_{1}}^{k_{1}} C_{N_{2}}^{k_{2}}} \nonumber \\
& \times \exp \left[\frac { \alpha } { \sqrt { 2 } } \left(e^{-i \tau\left[2 k_{1}+2 k_{2}-\left(N_{1}+N_{2}\right)\right]} a_1^{\dagger}\right.\right. \nonumber \\
& \left.\left.+e^{i \tau\left[2 k_{1}+2 k_{2}-\left(N_{1}+N_{2}\right)\right]} a_2^{\dagger}\right)\right]\left|k_{1}, k_{2}\right\rangle|0\rangle
\end{align}

After the light pulse passes through two BECs, the beam splitter operation transforms the photonic operators from $a_1,a_2$ to $c,d$.

\begin{align}
c^{\dagger} & =\frac{1}{\sqrt{2}}\left(a_1^{ \dagger}+i a_2^{\dagger}\right) \\
d^{\dagger} & =-\frac{1}{\sqrt{2}}\left(i a_1^{ \dagger}+a_2^{\dagger}\right)
\label{transformation operators}
\end{align}
then it exhibits a more precise manifestation

\begin{align}
|\tilde{\Psi}(\tau)\rangle&= \frac{e^{-\frac{|\alpha|^{2}}{2}}}{\sqrt{2^{\left(N_{1}+N_{2}\right)}}} \sum_{k_{1}, k_{2}} \sqrt{C_{N_{1}}^{k_{1}} C_{N_{2}}^{k_{2}}}\left|k_{1}, k_{2}\right\rangle \nonumber \\
& \times \exp \left[-\alpha e^{\frac{i \pi}{4}} i \sin \left(M+\frac{\pi}{4}\right) c^{\dagger}\right.  \nonumber \\
& \left.+\alpha e^{\frac{i \pi}{4}} i \cos \left(M+\frac{\pi}{4}\right) d^{ \dagger}\right]|0\rangle \nonumber \\
&=\frac{1}{\sqrt{2^{\left(N_{1}+N_{2}\right)}}} \sum_{k_{1}, k_{2}} \sqrt{C_{N_{1}}^{k_{1}} C_{N_{2}}^{k_{2}}}\left|k_{1}, k_{2}\right\rangle  \nonumber \\
&\times\left|-\alpha e^{i\frac{\pi}{4}} i\sin \left(M+\frac{\pi}{4}\right) \right\rangle_{c}\left|\alpha e^{i\frac{\pi}{4}} i\cos \left(M+\frac{\pi}{4}\right)\right\rangle_{d}
\end{align}

where $M=[2k_1+2k_2-(N_1+N_2)]\tau$. Finally, using the photon projection operator $| n_c,n_d \rangle$, the state collapses to an entangled state,

\begin{align}
& |{\tilde{\psi}}_{n_c n_d}(\tau)\rangle =  \frac{1}{\sqrt{\mathcal{N}}} \frac{1}{\sqrt{2^{\left(N_{1}+N_{2}\right)}}} \times \frac{e^{-\frac{|\alpha|^{2}}{2}} \alpha^{n_{c}+n_{d}}}{\sqrt{n_{c} !} \sqrt{n_{d} !}} \nonumber \\
&\times \sum_{k_{1}, k_{2}} \sqrt{C_{N_{1}}^{k_{1}} C_{N_{2}}^{k_{2}}} \times \sin(M+\frac{\pi}{4})^{n_c}\cos(M+\frac{\pi}{4})^{n_d}\left|k_{1}, k_{2}\right\rangle
\label{eq:FinaleUnnormalized}
\end{align}
%


The normalization factor ${\cal N}$ in eq. $(\ref{eq:FinaleUnnormalized}) $ is given by,
\begin{align}
\mathcal{N}=\frac{1}{2^{\left(N_{1}+N_{2}\right)}} \sum_{k_{1}, k_{2}} C_{N_{1}}^{k_{1}} C_{N_{2}}^{k_{2}}  {(\frac{e^{-\frac{|\alpha|^{2}}{2}} \alpha^{n_{c}+n_{d}}}{\sqrt{n_{c} !} \sqrt{n_{d} !}})}^2 \nonumber \\
\sin(M+\frac{\pi}{4})^{2n_c}\cos(M+\frac{\pi}{4})^{2n_d}
\end{align}
The most likely photon counts concerning the state (\ref{eq:FinaleUnnormalized}) distribute around $n_{c} +n_{d}\approx {|\alpha|}^{2} $.  In the case of $ |\tau| \lesssim 1/\sqrt{N} $, $ N \gg 1 $, $N_p=n_c+n_d= |\alpha |^2 \gg 1, M \gg 1 $, the wave function can be approximated as \cite{AristizabalZuluaga2021QuantumNM}  
\begin{align}
 | \tilde{\psi}_{n_c n_d  }^\text{approx} ( \tau ) \rangle & \propto \left(\frac{2}{N \pi}\right)^{\frac{1}{2}}\sum_{k_{1}, k_{2}}  e^{-\frac{1}{4 N}\left[\left(2 k_{1}-N\right)^{2}+\left(2 k_{2}-N\right)^{2}\right]}
 \nonumber \\
& \times \exp \left[-8 N_{p} \tau^{2}\left(k_{1}+k_{2}-N\right)^{2}\right]
\label{approximatewave}
\end{align}

From this approximated state, it is obvious that the distribution is very similar to the two-mode squeezed state. In the limit of large $n_c\tau^2$, the approximated state (\ref{approximatewave}) could be further approximated as
\begin{align}
\left|\tilde{\psi}_{\lim }^{\text {approx }}\right\rangle \rightarrow\left(\frac{4}{\pi N}\right)^{1 / 4} \sum_{k=0}^{N} e^{-\frac{2}{N}\left(k-\frac{N}{2}\right)^{2}}|k\rangle|k\rangle
\end{align}
which shows some similarities to the spin-EPR state that possesses maximal entanglement.

\subsection{Photon probability distributions}
\label{photon_distribution chapter}

As is described in Ref.\cite{pettersson2017light}, the photon measurement outcomes $n_c,n_d$ may not be explicitly known. Naturally, different choices of $n_c,n_d$ will produce different entangled states. It is necessary to examine the photon probability distributions at various time scales which are calculated by

\begin{align}
\mathcal{P} =Tr \big[\rho_{0} (k_1,k_2,k_1^{'},k_2^{'})\big]
\label{photon_expression}
\end{align}

\begin{figure}[t]
\centering
\includegraphics[width=\columnwidth]{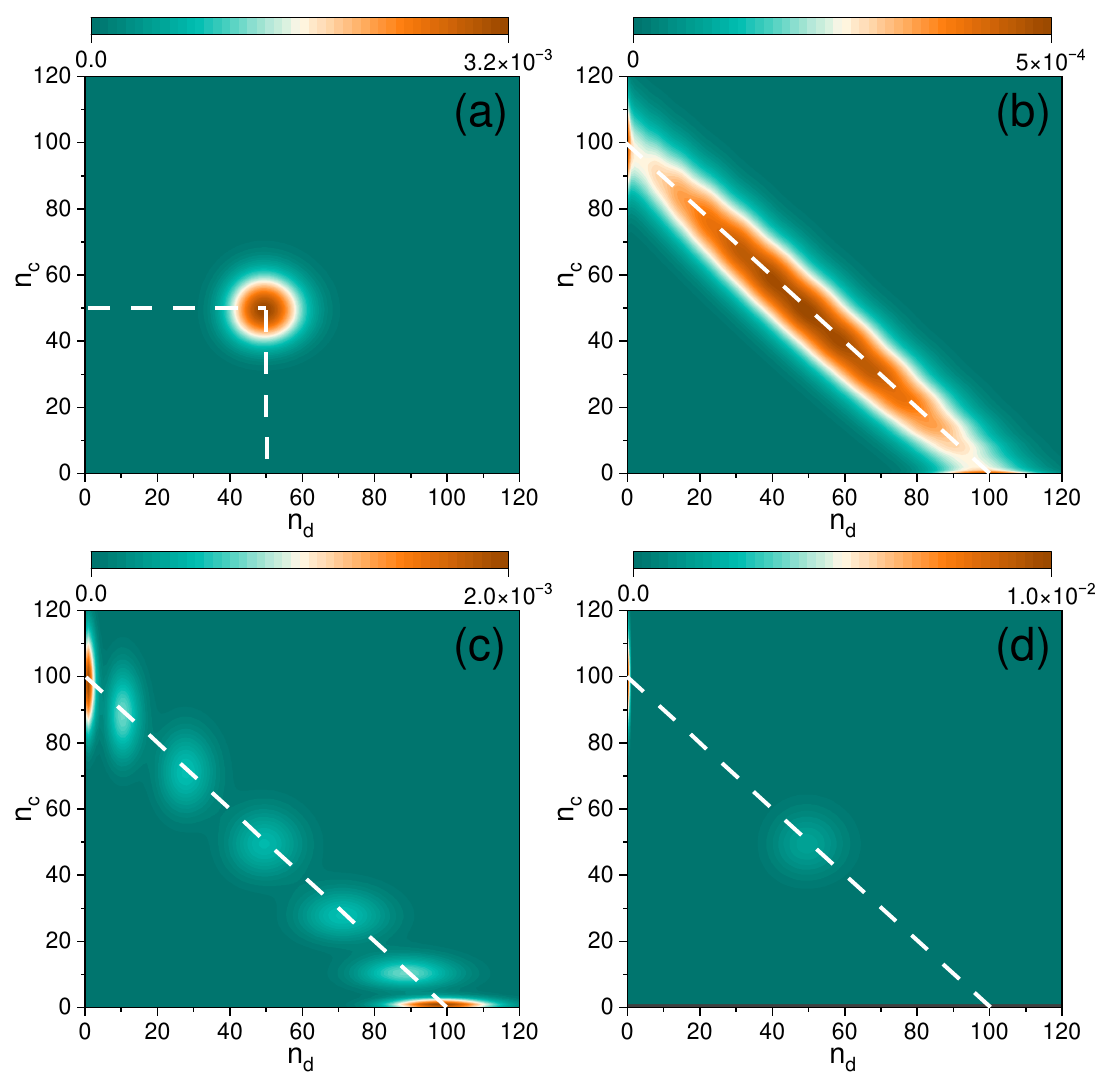}
\caption{Photon probability distribution (\ref{photon_expression})  without photon loss decoherence for (a) $\tau=0$. (b) $\tau=1/N$. (c) $\tau=1/(2\sqrt{N})$. (d) $\tau=\pi/8$. The dashed line represents $n_c+n_d={|\alpha|}^2$ which has the most obvious range of observation. Common parameters used are $N=20,\alpha=10$.}
\label{photon distribution}
\end{figure}

Fig.\ref{photon distribution} displays the photon probability distributions for different times. In Fig.\ref{photon distribution}(a), the photon probability distribution takes the form of a circular distribution, with the center at $n_c=n_d=50$ for $\tau=0$, which is a typical coherent light field distribution pattern. For $\tau=1/N$, as shown in Fig.\ref{photon distribution}(b), the photon probability distribution is approximately elliptical with the axes roughly along $n_c+n_d={|\alpha|}^2$. In Fig.\ref{photon distribution}(c), for $\tau=1/(2\sqrt{N})$ that beyond the Holstein-Primakoff (HP) approximation, photons measured in $c,d$ modes $(n_c,n_d)$ are mainly distributed in $(100,0)$ and $(0,100)$, with sporadic distributions occurring in between, taking the form of a rough bridge. Concerning $\tau=\pi/8$, the photons are mainly concentrated in $(0,100)$, indicating that the photons tend to be distributed more in the $c$ optical mode over a more extended period. Based on the photon probability distribution, we will focus on photon measurement outcomes in the upcoming research: $n_d=n_c=50$, $n_d=60, n_c=40$ and $n_d=70, n_c=30$.

\section{Photon loss channel } \label{photon_loss_chap}

\subsection{Density matrix}

The conventional approach to solving the quantum master equation involves using the phase space representation  \cite{wolfgang2001quantum}, such as the Wigner function or Glauber-Sudarshan function, to transform the equation into the Fokker-Planck equation or Langevin equation. Another method, known as the super-operator method \cite{arevalo1998quantum}, has limitations due to its lack of eigenstates. These limitations were discovered by Arealo-Aguilar and Moya-Cessa when dealing with arbitrary initial states. To overcome these limitations, a new method was developed using the representation of thermally entangled states \cite{PhysRevA.80.022115,fan2008mod}. This method maps the density operator to a vector in a two-mode Fock space, where one mode represents the system and the other represents the imaginary mode. By performing this action, the main equation takes on a form similar to the Schrödinger time evolution equation equation.

In the interaction representation, the photon loss channel could be given by 

\begin{align}
\frac{d \rho}{d t}=(-i)^{2} \gamma\left[a^{\dagger} a,\left[a^{\dagger} a, \rho\right]\right]+\chi \left(2 a \rho a^{\dagger}-a^{\dagger} a \rho-\rho a^{\dagger} a\right)
\label{master_equation}
\end{align}
In equation \ref{master_equation}, the first term denotes the phase-damping process, while the subsequent term signifies the process of amplitude attenuation. The parameters $\gamma$ and $\chi$ correspond to the phase damping and amplitude attenuation coefficients. The QND interaction rate is represented by the parameter $q$, while the loss rate is denoted by the parameter $\bar{\chi}$. The loss and QND interaction rates can be adjusted by $\bar{\chi}=\hbar \chi /q$.

The time evolution of the density matrix for the system is calculated as follows

\begin{align}
\rho(t) & =\sum_{l, m=0}^{\infty} \frac{\left(1-e^{-2 \chi t}\right)^{l}}{l ! m !}\left(2 \gamma t a^{\dagger} a\right)^{m} \nonumber \\
& \times e^{-\gamma t\left(a^{\dagger} a\right)^{2}-\chi t a^{\dagger} a}  a^{l} \rho_{0} a^{\dagger l} e^{-\gamma t\left(a^{\dagger} a\right)^{2}-\chi t a^{\dagger} a}\left(a^{\dagger} a\right)^{m} \nonumber \\
& = \sum_{l, m=0}^{\infty} M_{l, m} \rho_{0} M_{l, m}^{\dagger}
\end{align}

and for the system of two BECs, it takes the following form,
\begin{align}
\rho_{\text{atom}}&(k_1,k_2,k_1^{'},k_2^{'}) =  \langle n_c | \langle n_d | \sum_{i,j,l,p=0}^{\infty} M_{i, j}^{\text{(1)}}  (t)  M_{l, p}^{\text{(2)}}  (t) \nonumber \\
& \times \rho_{0}(k_1,k_2,k_1^{'},k_2^{'}) {M_{l, p}^{\text{(2)}}}^{\dagger} (t) 
 {M_{i, j}^{\text{(1)}}}^{\dagger} (t) | n_c \rangle | n_d \rangle \nonumber \\ ,
& = \rho_{0}(k_1,k_2,k_1^{'},k_2^{'})\mathcal{L}[(2k_1+2k_2-2k_1^{'}-2k_2^{'})\tau]
\label{final density in Krauss-like}
\end{align}

where $\rho_{0}(k_1,k_2,k_1^{'},k_2^{'})$ is the initial density matrix of the system, in our case, it is 
\begin{align}
\rho_{0} (k_1,k_2,k_1^{'},k_2^{'})&= |{\tilde{\psi}}_{n_c n_d}(\tau)\rangle \langle {\tilde{\psi}}_{n_c n_d}(\tau) | ,
\label{initial density matrix}
\end{align}   

The $\mathcal{L}$ function which includes decoherence effects is given by $\mathcal{L}(\upsilon)=e ^{(n_c+n_d)} (-2\chi t) \exp[(1-\exp^{-2\chi t}){|\alpha|}^2 \cos(\upsilon)]$. The Kraus operator that involves both phase damping and amplitude attenuation is referred to here.
\begin{align}
M_{l, m}=\sqrt{\frac{\left(1-e^{-2 \chi t}\right)^{l}}{l ! m !}(2 \gamma t)^{m}}\left(a^{\dagger} a\right)^{m} e^{-\gamma t\left(a^{\dagger} a\right)^{2}-\chi t a^{\dagger} a} a^{l} .
\end{align}

Although the Krauss operator includes a phase-damping coefficient, our calculation shows that it does not affect the entangled state generated in this paper. Therefore, our study in this paper primarily focuses on the amplitude attenuation effects of the second term in the master equation. The relationship between the number of photons remaining in the system and the amplitude attenuation coefficient due to photon loss is adjusted by $n_{\text{left}}=e^{-2\chi t}{|\alpha|}^2$.

\subsection{Wigner distributions }

The previous analysis conducted by Pettersson \cite{pettersson2017light}  provides a preliminary understanding of the entangled state described by equation (\ref{eq:FinaleUnnormalized}). It is customary to compute the Wigner functions to explore more properties of such states.

The Wigner function \cite{PhysRev.40.749,PhysRevA.49.4101,Schmied_2011} is represented by 

\begin{align}
W(\theta, \varphi)=\sum_{l=0}^{2 j} \sum_{q=-l}^{l} \rho_{l q} Y_{l q}(\theta, \varphi)
\label{wigner expression}
\end{align}

where $j=\frac{N}{2}$, $Y_{l q}$ are the spherical harmonics function and $\rho_{l q}$ is defined as

\begin{align} 
\rho_{l q}&=\sum_{m_{1}=-j}^{j} \sum_{m_{2}=-j}^{j}(-1)^{j-m_{1}-q}\left\langle j m_{1} ; j-m_{2}\mid l q \right\rangle \\ \nonumber &\times
\left\langle j m_{1}|\rho| j m_{2}\right\rangle.
\end{align}
where $\left\langle j m_{1} ; j-m_{2}\mid l q \right\rangle $ is the Clebsh-Gordan coefficient. $\left\langle j m\right\rangle$ is a Dicke state which could be written as 

\begin{align}
|j m\rangle=|k=j+m\rangle_{N=2j}=\frac{\left(a^{\dagger}\right)^{j+m}\left(b^{\dagger}\right)^{j-m}}{\sqrt{(j+m) !(j-m) !}}|0\rangle,
\end{align}

The Wigner function denotes the quasi-probability distribution of a quantum state on the Bloch sphere. Due to the presence of two Bose-Einstein condensates (BECs), each with two components, the Wigner function possesses four degrees of freedom ($\theta_{1}, \phi_{1}, \theta_{2}, \phi_{2}$). Therefore, visualizing the Wigner function directly is quite a challenging task. The marginal and conditional Wigner functions are then developed to solve this problem. The difference between these two calculations is that the marginal Wigner function is calculated by tracing over one BEC. However, we perform a projection $\left | k \right \rangle\left \langle  k \right |$ on the Fock basis in calculating the conditional Wigner function. The conditional Wigner function is characterized as

\begin{align}
&W_{c}(\theta_1, \phi_1)=\sum_{l=0}^{N} \sum_{k_{1},
 k_{1}^{\prime} =0}^{N}
(-1)^{N-2 k_{1}+k_{1}^{\prime}} \rho(k_1,k_2,k_1^{'},k_2) 
 \nonumber \\  & \times \left\langle\frac{N}{2}, 
k_{1}-\frac{N}{2} ; \frac{N}{2}, \frac{N}{2}-k_{1}^{\prime} 
\mid l, k_{1}-k_{1}^{\prime}\right\rangle  Y_{l, k_1-k_{1}^{\prime}}\left 
( \theta_1,\phi_1  \right )  
\end{align}

\begin{figure}[t]
\includegraphics[width=\columnwidth]{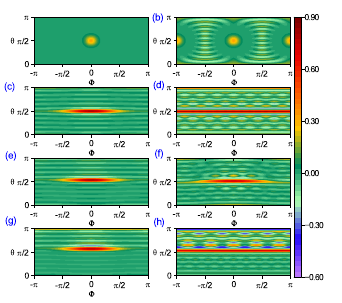}
\caption{\label{Conditional wigner distribution}Conditional Wigner distributions for the initial state (\ref{eq:FinaleUnnormalized}) with $k_2=N/2$. (a) $\tau=0$ for all three cases. (b) $\tau=\pi/8$ for all three cases. (c-d) $n_d=n_c=50$, with the corresponding interaction times $\tau=1/N,1/(2\sqrt{N})$ from left to right. (e-f) $n_d=60,n_c=40$, with $\tau=1/N,1/(2\sqrt{N})$ from left to right. (g-h) $n_d=70,n_c=30$ for $\tau=1/N,1/(2\sqrt{N})$, respectively. In all plots, the total atom numbers are $N=20$, and the amplitude of coherent light is set as $\alpha=10$. }
\end{figure}

Fig.\ref{Conditional wigner distribution} shows the conditional Wigner distributions in decoherence-free cases. The density matrix $\rho(k_1, k_2, k_1^{'}, k_2)$ corresponds to the initial density matrix $\rho_{0}(k_1, k_2, k_1^{'}, k_2)$ described in eq.(\ref{initial density matrix}). On the other hand, the other plots use the density matrix $\rho_{\text{atom}}(k_1, k_2, k_1^{'}, k_2)$ defined in eq.(\ref{final density in Krauss-like}). We consider the conditional Wigner distributions with a $k = N/2$ projection.  For the initial state (\ref{initial density matrix}) without any decoherence, the Wigner function shows a Gaussian distribution which is a spin coherent state centered at $\theta=\pi/2,\phi=0$ for $\tau=0$ in all three cases. At $\tau=\pi/2$, distinct fringes appear that connect the two peaks, corresponding to the \text{Schrödinger} cat state $|\pi/2,0\rangle\rangle|\pi/2,0\rangle\rangle+|\pi/2,\pi\rangle\rangle|\pi/2,\pi\rangle\rangle$ regardless of the value of photons $n_c$. When $\tau=1/N$, the Wigner distribution evolves into an ellipse shape along the $z$ direction due to a squeezing effect, another important feature at this time scale is that the value of the Wigner function becomes smaller as $n_d$ increases, indicating that an increase in $n_d$ leads to a more non-Gaussian entangled state. For a duration of time $\tau=1/(2\sqrt{N})$ in Fig.\ref{Conditional wigner distribution}(d-f), distribution along the equator emerges, which is a characteristic feature of a non-Gaussian state. Notably, the negative portion of the distribution increases as time progresses, indicating the highly non-classical nature of the state. For $n_d=50$, the Wigner function is mainly distributed on the equator, whereas for $n_d=60$, the Wigner function is mainly distributed in an elliptical form. With the further increase in $n_d$, the Wigner distribution is concentrated near the equator, but some wave spots appear on the south pole of the Bloch sphere.

\begin{figure}[t]
\includegraphics[width=\columnwidth]{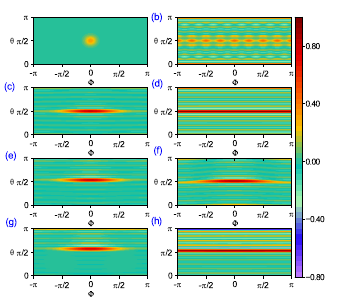}
\caption{\label{wigner_decoherence}Conditional Wigner distributions for the initial state (\ref{eq:FinaleUnnormalized}) under photon loss channel with $k_2=N/2$. (a) $\tau=0$ for all three cases. (b) $\tau=\pi/8$ for all three cases. (c-d) $n_d=50,n_c=50$. (e-f) $n_d=60,n_c=40$. (g-h) $n_d=70,n_c=30$. For plots (c-h), the corresponding interaction times are $\tau=1/N,1/(2\sqrt{N})$ from left to right. In all plots, the total atom numbers are $N=20$, the amplitude attenuation coefficient is $\bar{\chi}=\hbar \chi/q=0.3$, and the amplitude of coherent light is chosen as $\alpha=10$. }
\end{figure}

The figures shown in Fig.\ref{wigner_decoherence} account for the decoherence effect caused by photon loss that mainly results in amplitude attenuation. It has been observed that in the time region of $\tau \sim 1 / N$, where the HP approximation is valid, there is no alteration in the Wigner distribution for all three cases $n_d=50,60,70$, indicating its robustness during this time. However, at $\tau=1/(2\sqrt{N})$, the previous spot distribution is smoothed out, which suggests that the entangled state is affected by decoherence due to photon loss. In Fig.\ref{wigner_decoherence}(f), it can be seen that the \text{Schrödinger} cat state is significantly disrupted, and lantern-like distributions emerge along the equator. This further demonstrates the susceptibility of the \text{Schrödinger} cat state to decoherence.

\subsubsection{Entanglement and Fidelity}
\label{entanglent section}

To quantify the impact of photon loss on the entanglement between BECs, we employ logarithmic negativity as a measure, expressed by the following mathematical formula \cite{PhysRevLett.95.090503,PhysRevA.65.032314}.

\begin{align}
E=\log _{2}\left\|\rho^{T_{2}}\right\|
\label{logarithmic negativity}
\end{align}
here, $\rho^{T_2}$ is partial transpose on the second BEC. The entanglement entropy $E$ is normalized according to the maximum entanglement

\begin{align}
E_{\max }=\log _{2}(N+1)
\end{align}

Fidelity is determined by calculating the overlap between the analyzed and target state. The spin EPR state exhibits the highest level of entanglement between the two BECs and is widely used in quantum teleportation masks \cite{PhysRevA.91.063832,zhao_enhancing_2023}. Fidelity is specifically defined in the following form 

\begin{align}
F=\left\langle \mathrm{EPR}\right|\rho_{\text{atom}}(k_1,k_2,k_1^{'},k_2^{'})\left| \mathrm{EPR}\right\rangle
\label{fidelity}
\end{align}
and the EPR state \cite{PhysRev.47.777} is invariant in any basis 

\begin{align}
\left|\mathrm{EPR}\right\rangle=\frac{1}{\sqrt{N+1}} \sum_{k=0}^{N}|k\rangle|k\rangle  
\label{EPR}
\end{align}
J

\begin{figure}[t]
\centering
\includegraphics[width=\columnwidth]{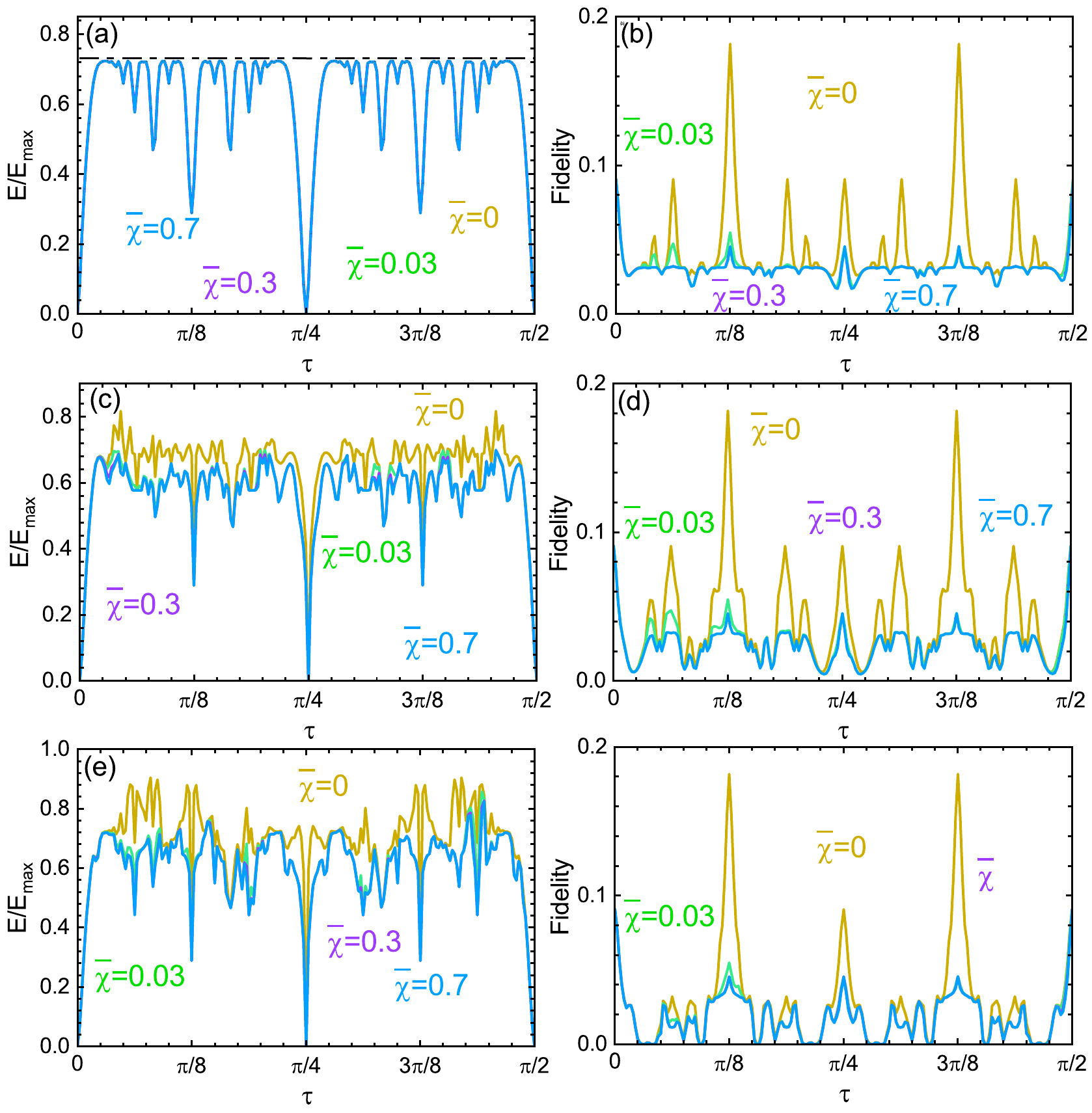}
\caption{Entanglement and fidelity for the state (\ref{eq:FinaleUnnormalized}) in the presence of photon loss channel. (a) Entanglement under photon loss channel vs. interaction times with $\tau$  as marked. The photon measurement outcomes are $n_d=n_c=50$. The dashed line in this plot indicates the ceiling of entanglement values. (b) Fidelity under photon loss channel vs. interaction times  $\tau$ with photon outcomes $n_d=n_c=50$. (c) Entanglement under photon loss channel vs. interaction times  $\tau$. The corresponding photon numbers are set as $n_d=60,n_c=40$. (d) Fidelity under photon loss channel vs. interaction times $\tau$ for $n_d=60,n_c=40$. (e-f) Entanglement and fidelity under photon loss channel vs. interaction times $\tau$ from left to right. The photon measurement outcomes are $n_d=70,n_c=30$. The amplitude attenuation coefficients are in units of $q/\hbar$, and the time in units $ \hbar/q $. }
\label{entanglment plots}
\end{figure}

In Fig.\ref{entanglment plots}, we present the entanglement and fidelity for the state (\ref{eq:FinaleUnnormalized}) in the presence of the photon loss channel. The entanglement and fidelity values are plotted against the time variable $\tau$. The entanglement exhibits a devil's staircase structure when $n_d=n_c=50$, as shown in Fig.\ref{entanglment plots}(a). The maximum entanglement value obtained in this case is $E/E_{max}=0.73$, which is similar to the results reported in Refs.\cite{PhysRevA.88.023609,shuaiaqs,AristizabalZuluaga2021QuantumNM}. It is worth noting that the period of the devil's staircase is $\pi/4$ in this case. This is originated from the fact when $n_d=n_c$, the trigonometric functions in the wave functions (\ref{eq:FinaleUnnormalized}) follow $\sin(M+\pi/4)^{n_d} \cos(M+\pi/4)^{n_d}=(1/2\cos(2M))^{n_d}$. For a more general case, as shown in Fig.\ref{entanglment plots}(c,e), the devil's staircase exhibits a period of $\pi/2$. As the amplitude attenuation coefficient increases, we observe the entanglement remains constant throughout the time scale in Fig.\ref{entanglment plots}(a). Now turning to Fig.\ref{entanglment plots}(c,e), where the $d$ mode photon outcomes $n_d=60,70$ respectively, the entanglement without photon loss exhibits exotic properties, with the entanglement values $E/E_{max}$ reaching above 0.9 at some points. We attribute this to the generation of a more non-Gaussian state. The increase in entanglement greatly enhances the potential application of entangled states in quantum teleportation tasks. Additionally, the entanglement decreases significantly with decoherence increasing, but the entanglement value remains a considerable value.

In Fig.\ref{entanglment plots}(b,d,f), we plot the fidelity of the EPR state acting as the reference state. There are two noticeable differences when compared to entanglement plots. The fidelity remains relatively low at $\bar{\chi}=0.7$ while the entanglement value exceeds 0.73. Secondly, at $\tau=0,\pi/2$, the entanglement value remains at zero, but the fidelity value is the contrary. This is because the ensemble comprises two spin coherent states at these time points, where the fidelity is $1/(N+1)$. The amplitude coefficient of the spin coherent state (\ref{coherent state expression}) is not time-dependent, making the fidelity relatively robust even when photon loss effects are considered. In Fig.\ref{entanglment plots}(d), it is evident that the relationship between fidelity and amplitude attenuation coefficients at different times follows a similar pattern to the entanglement plots in Fig.\ref{entanglment plots}(a). For the decoherence-free case, the highest fidelity is observed at $\tau=\pi/8,3\pi/8$ when they are Schrödinger cat states $|\pi/2,0\rangle\rangle|\pi/2,0\rangle\rangle+|\pi/2,\pi\rangle\rangle|\pi/2,\pi\rangle\rangle$. When compared to the $n_d=50$ case in Fig.\ref{entanglment plots}(d), the fidelity values in Fig.\ref{entanglment plots}(e,f) where $n_d=60,70$ have significantly decreased. As the amplitude attenuation coefficient $\bar{\chi}$ increases, the fidelity of the three cases $n_d=50,60,70$ significantly declines owing to stronger photon loss effects. In Fig.\ref{entanglment plots}(f), the fidelity has even become zero at some time scales. This means that larger $n_d$ produces entangled states that are more unstable than the $n_d=50$ case.


\subsubsection{Probability distribution}
\label{probability distribution section}

We also consider the probability distribution of measuring  state with decoherence in various bases |$k_1, k_2\rangle^{(l,m)}$, where $l,m\in \left \{ x,y,z \right \} $ (Appendix.\ref{Fock state elements})
\begin{align}
p_{l,m}(k_1,k_2)=\langle k_1, k_2 |^{(l,m)} \rho_{\text{atom}}(k_1,k_2,k_1^{'},k_2^{'}) |k_1, k_2\rangle^{(l,m)}. \label{correlation}
\end{align}
and the notation $| k_1, k_2 \rangle^{(l,m)}$ is the Fock state in the basis $l,m$ 
\begin{align}
| k_1, k_2 \rangle^{(l,m)} = | k_1 \rangle^{(l)} \otimes  | k_2 \rangle^{(m)}
\end{align}

The relationship between the Fock operator ${S}_{i}^{(O)},i \in \left \{ 1,2 \right \},O \in \left \{ x,y,z \right \} $ and the Fock states for different bases is obvious from the following expressions

\begin{align}
{S}_{i}^{(O)}|k\rangle^{({O})}=(2 k-N)|k\rangle^{({O})}
\end{align}

\begin{figure}[t]
\centering
\includegraphics[width=\columnwidth]{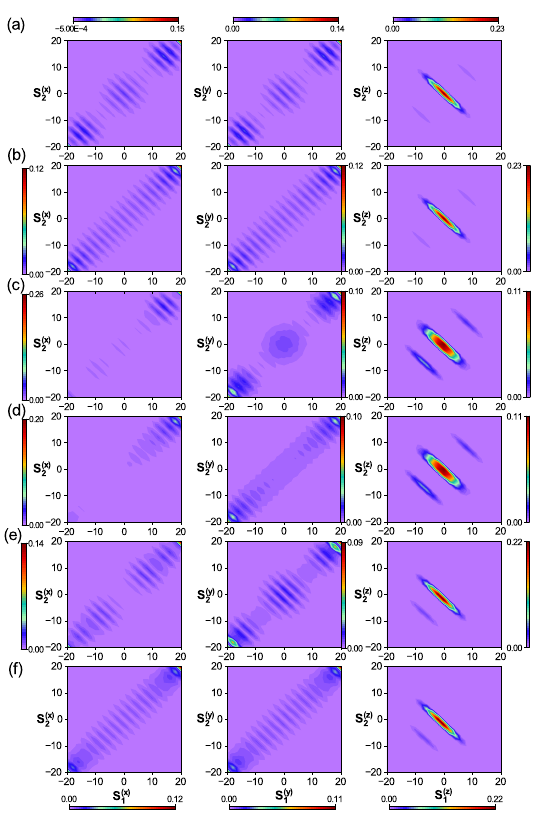}
\caption{Probability distributions under photon loss channel in various bases, as defined in (\ref{correlation}). (a) Without any photon loss for $n_d=50,n_c=50$. (b) With photon loss decoherence for $n_d=50,n_c=50$. (c-d) The decoherence-free and decoherence cases for $n_d=60,n_c=40$. (e-f) The decoherence-free and decoherence cases for $n_d=70,n_c=30$. The parameters used are $N=10,\tau=0.1$, and $\bar{\chi}=\hbar\chi/q=0.3$ for all decoherence cases. }
\label{density2atom}
\end{figure}

Fig.\ref{density2atom} illustrates the probability distribution of the state (\ref{eq:FinaleUnnormalized}) with photon loss decoherence. We observe the correlations in the $S_x$ spin basis for the decoherence-free case in Fig.\ref{density2atom}(a). The primary distribution is concentrated around the values of $S_1^x= N,S_2^x= N$ and $S_1^x= -N,S_2^x= -N$, consistent with the initial spin coherent state. The distribution in the $S_y$ spin basis is very similar to that in the $S_x$ spin basis, exhibiting a correlation in a diagonal direction. As for the $S_z$ spin basis, the primary correlation peak appears in the direction $S_1^z+S_2^z=N$. Besides, weak secondary peaks are apparent on two sides of the primary correlation peak, resulting from the periodicity of triangle functions as stated in Ref.\cite{pettersson2017light}. For the $n_d=60,n_c=40$ decoherence-free case in Fig.\ref{density2atom}(c), there are two main changes compared with $n_d=n_c=50$ decoherence-free case. Firstly, the correlations in the $S_x $ spin basis mainly distributed in $S_1^x= N,S_2^x= N$ while  $S_1^x= N,S_2^x= N$ and $S_1^x= -N,S_2^x= -N$ in Fig.\ref{density2atom}(a). Secondly, the anti-correlation primary peak in the $S_z$ spin basis is broadened, and two secondary peaks are more obvious than that in the $n_d=n_c=50$ case. Fig.\ref{density2atom}(e) shows probability distributions of the state with photon outcomes $n_d=70,n_c=30$, the secondary peaks only appear on one side in the $S_z$ spin basis, indicating more exotic properties.

When the photon loss channel is introduced in Fig.\ref{density2atom}(b,d,f), the probability distributions are different from that in the decoherence-free case. The correlations in the $S_x$ spin basis are decreased while increasing in the $S_y$ spin basis. This type of dephasing can be seen as a random rotation about the $S_z$ basis on both BECs, resulting in approaching the $S_y$ basis.  The anti-correlations in the $S_z$ spin basis in all three cases are unvaried, this is because the wave function amplitude tends to be maximum when $k_1+k_2-N=0$ and the same case for $k_1^{'}+k_2^{'}-N=0$, so the $\mathcal{L}(\upsilon)$ decoherence term tends to be unity all the time.

\subsubsection{Expectation values and variances}
\label{variance section}

\begin{figure}[t]
\centering
\includegraphics[width=\columnwidth]{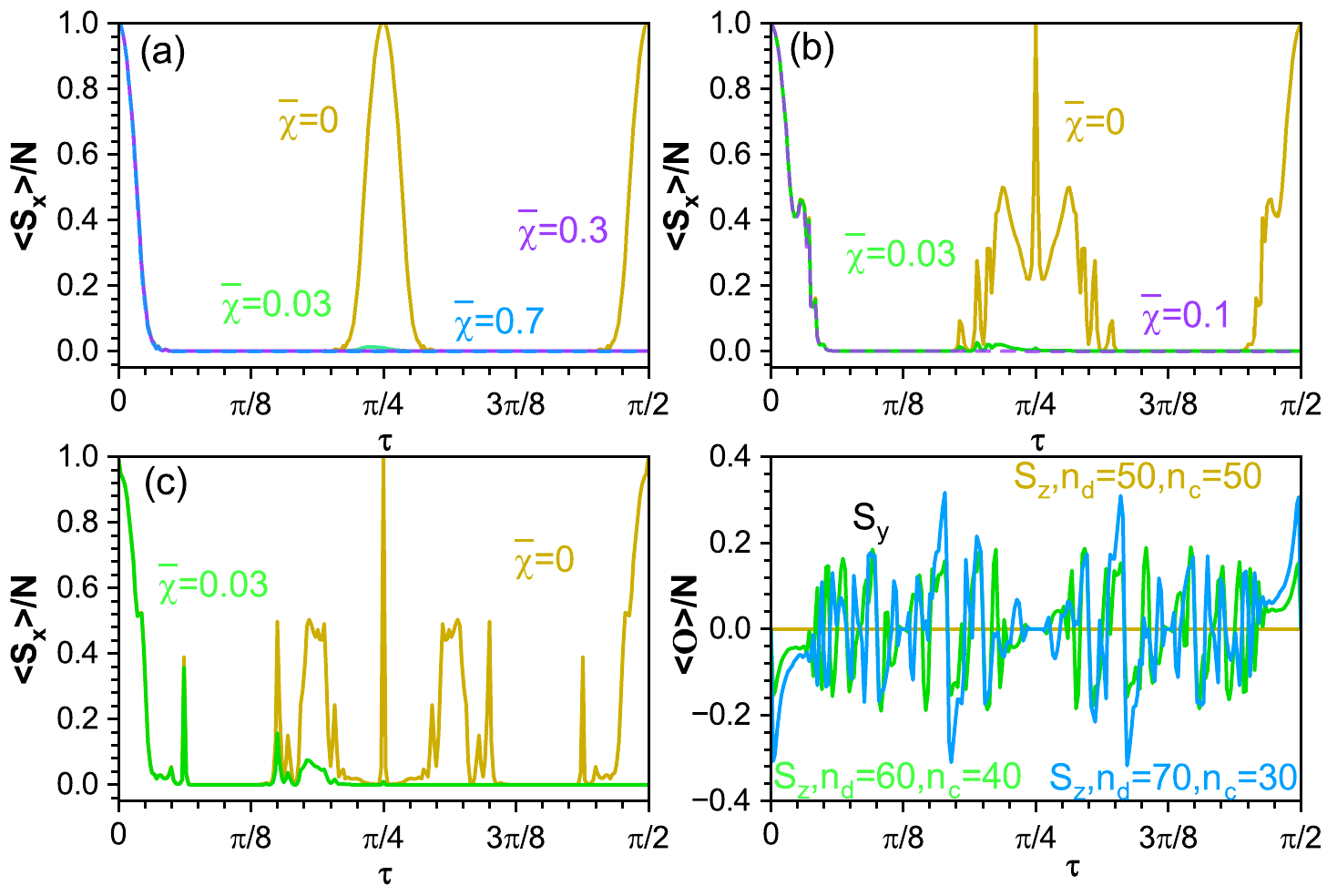}
\caption{The time evolution of expectation values for the state (\ref{eq:FinaleUnnormalized}) with photon loss decoherence. (a) Expectation value $\left \langle S_x \right \rangle$ with $n_d=n_c=50$. (b) Expectation value $\left \langle S_x \right \rangle$ for $n_d=60,n_c=40$. (c) Expectation value $\left \langle S_x \right \rangle$, $n_d=70,n_c=30$. (d) Expectations for the operators $\left \langle S_y,S_z \right \rangle$ for $n_d=50,60,70$ respectively. 
The parameters used are $N=20,\alpha=\sqrt{n_c+n_d}=10$. For the $n_d=n_c=50$ case, the amplitude attenuation coefficient is chosen as $\bar{\chi}=\hbar \chi/q=0,0.03,0.3,0.7$. For photon measurement outcomes $n_d=60,n_c=40$, the amplitude attenuation coefficient is $\bar{\chi}=\hbar \chi/q=0,0.03,0.1$ while $\bar{\chi}=\hbar \chi/q=0,0.03$ for $n_d=70,n_c=30$. }
\label{expectation_plot}
\end{figure}

The expectation values for an operator $\Lambda$ could be written as $\left \langle \Lambda \right \rangle =\text{Tr} (\Lambda \rho)$, and the corresponding variance is calculated by $\Delta=\left \langle \Lambda ^2 \right \rangle -{\left \langle \Lambda  \right \rangle} ^2$. In Fig.\ref{expectation_plot}, we present the expectation values of the state (\ref{eq:FinaleUnnormalized}) under photon loss decoherence. For the $n_d=n_c=50$ case in Fig.\ref{expectation_plot}(a), the expectation values $\left \langle S_x \right \rangle$ without decoherence is very similar to that in Refs.\cite{Gao_2022,AristizabalZuluaga2021QuantumNM,shuaiaqs}. $\left \langle S_x \right \rangle$ for $n_d=60,70$ case without decoherence in Fig.\ref{expectation_plot}(b,c) shows more oscillation peaks in the neighbourhood of $\tau=\pi/4$. The $n_d=n_c=50$ case with photon loss in Fig.\ref{expectation_plot}(a) shows a noticeable difference compared to the decoherence-free case. When $\bar{\chi}=0.03$, the expectation values of $S_x$ in the neighborhood of $\tau=\pi/4$ have been dramatically reduced. And it tends to be zero when $\tau \ge  37\pi/80$ which is very different from that in Ref.\cite{Gao_2022} where it tends to go back to unity. With the amplitude attenuation coefficient $\bar{\chi}=0.03$ further increasing, $\left \langle S_x \right \rangle$ remains at zero after it initially drops precipitously. We find the expectation value $\left \langle S_x \right \rangle$ at a larger time scale is more sensitive to photon loss because there are fewer photons over longer time scales according to $n_{\text{left}}=e^{-2 \bar{\chi} \tau }{|\alpha|}^2$. It also confirms the similar effects of photon loss in Fig.\ref{expectation_plot}(b,c) where the expectation values are largely reduced over a longer time duration. In  Fig.\ref{expectation_plot}(d), the expectation values of $\left \langle S_y \right \rangle$ and $\left \langle S_z \right \rangle$ are depicted as a function of interaction times in the presence of photon loss. The mean spin expectation value of $S_y$ stays firmly at zero, indicating that the mean spin is situated in the $x-z$ plane. The expectation values of $\left \langle S_z \right \rangle$ exhibit no variation, consistent with observations from the probability distribution. A positive $\left \langle S_z \right \rangle$ signifies that the mean spin is located in the northern hemisphere of the Bloch sphere, while a negative value indicates placement in the southern hemisphere. With an increase in  $d$ mode photon numbers $n_d$, we observe that the deviation of the expected values of $S_z$ from the center also increases, indicating a rise in variance.

\begin{figure}[t]
\centering
\includegraphics[width=\columnwidth]{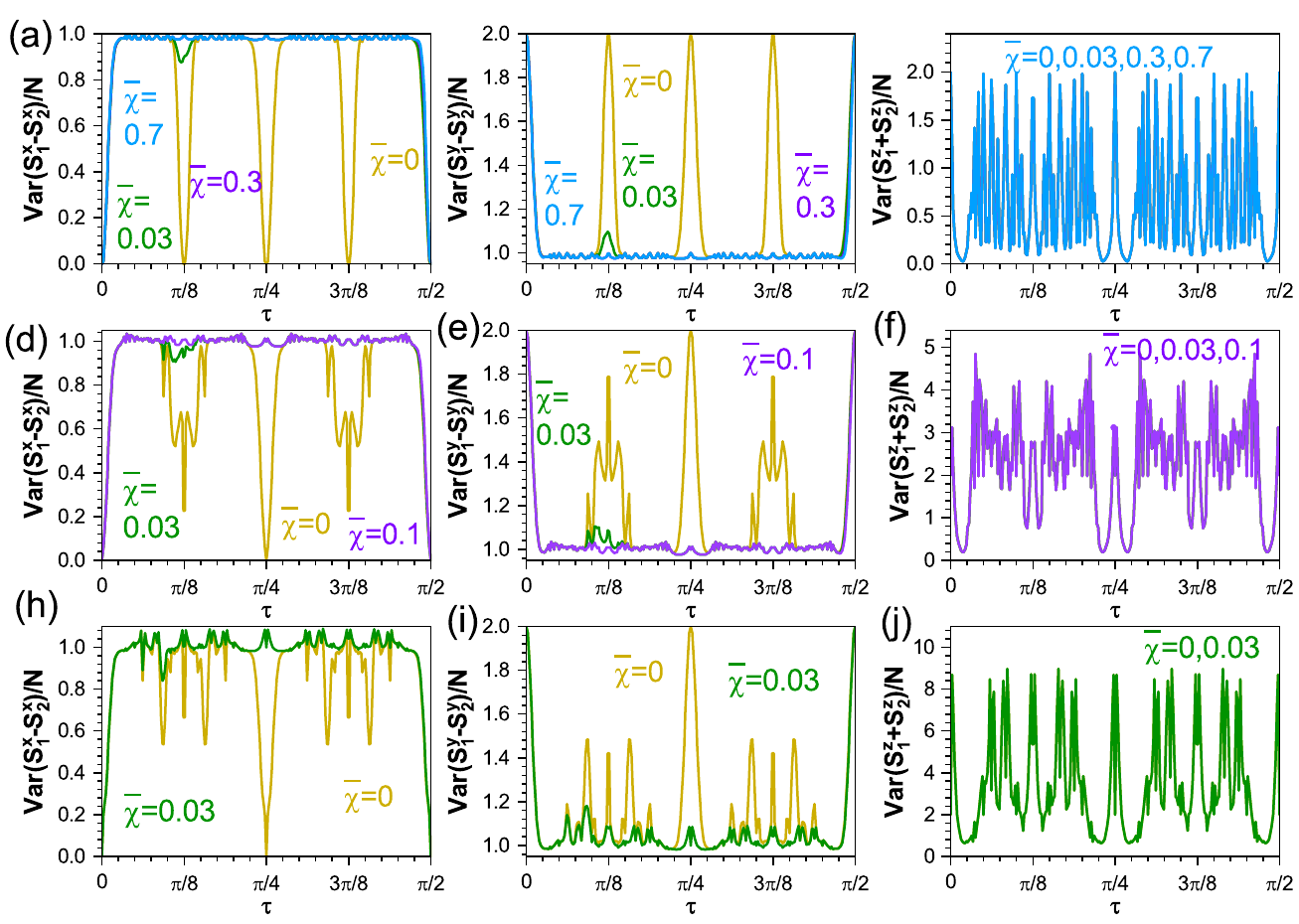}
\caption{The variances $\text{Var} (S_1^x-S_2^x),\text{Var} (S_1^y-S_2^y),\text{Var} (S_1^z+S_2^z)$ of the state (\ref{eq:FinaleUnnormalized}) under photon loss channel for different amplitude attenuation coefficients. (a-c) $n_d=n_c=50$; (d-f) $n_d=60,n_c=40$. (h-j) $n_d=70,n_c=30$. The common parameters used are $N=20,\alpha=\sqrt{n_c+n_d}=10$. For the $n_d=n_c=50$ case, the amplitude attenuation coefficient is chosen as $\bar{\chi}=\hbar \chi/q=0,0.03,0.3,0.7$. The amplitude attenuation coefficient is $\bar{\chi}=\hbar \chi/q=0,0.03,0.1$  regarding $n_d=60,n_c=40$ photon measurement outcomes, while $\bar{\chi}=\hbar \chi/q=0,0.03$ for the $n_d=70,n_c=30$ case.  }
\label{variance_all}
\end{figure}

The variances $\text{Var} (S_1^x-S_2^x),\text{Var} (S_1^y-S_2^y),\text{Var} (S_1^z+S_2^z)$ of the state (\ref{eq:FinaleUnnormalized}) under the photon loss channel for different amplitude attenuation coefficients are shown in Fig.\ref{variance_all}.
We first consider loss-free cases with photon measurement outcomes $n_d=50,60,70$. We find variances $\text{Var} (S_1^x-S_2^x),\text{Var} (S_1^y-S_2^y),\text{Var} (S_1^z+S_2^z)$ with photon measurement outcomes $n_d=n_c=50$  agrees well with that in Ref.\cite{Gao_2022} where we interpret the photon loss as an effective $S_z$ dephasing. It is found that for $n_d=60,70$ cases, all the three variances at $\tau=0,\pi/4,\pi/2$ show no differences with that in $n_d=50$ case which resulted from the state evolving to initial spin coherent state (\ref{becinitial}). With larger photon loss decoherence considered, more peaks appear in the neighborhood of $\tau=\pi/4$, showing highly non-Gaussian properties. Upon considering photon losses for amplitude attenuation coefficients $\bar{\chi}=0.03$ in Fig.\ref{variance_all}(a), it is observed that $\text{Var} (S_1^x-S_2^x)$ and $\text{Var} (S_1^y-S_2^y)$ are the same as in the absence of photon losses, except for the retention of appreciable values at $\tau=\pi/8$. This is due to the \text{Schrödinger} cat state near $\tau=\pi/8$, which is very sensitive to decoherence. The structure of the peaks tends to be flattened under larger photon loss effects. In general, the photon loss decoherence increases the variances of the operator $(S_1^x-S_2^x)$ while degrades $\text{Var} (S_1^y-S_2^y)$ on the contrary. The variances of the $(S_1^z+S_2^z)$ operators exhibit no disparity compared to scenarios without decoherence, attributed to the tendency of the $(S_1^z+S_2^z)$ operator to couple $2(k_1+k_2-N)$ terms, maintaining decoherence terms at unity in the density matrix. As expected, we witness the increase of $\text{Var} (S_1^z+S_2^z)$ with photon numbers in $d$ mode $n_d$ increasing.

\subsubsection{Correlation-based entanglement criteria }
\label{atom entanglement criteria}

Using logarithmic negativity as a measure to evaluate entanglement has been well-established. However, the practical application of this method is limited by the requirement for complete density matrix tomography. Fock state measurements, particularly valuable in the context of Bose-Einstein condensates, are constrained in their resolution utility on atom-number measurements. Compared with the methods above, low-order spin expectation values are the preferred observables in the experimental scenario.

Hofmann-Takeuchi criterion \cite{PhysRevA.68.032103} is applicable for any separable state, and it provides a convincing way of detecting correlations. This criterion does not include any expectation values of
spins $\left \langle S^x,S^y,S^z \right \rangle $, avoiding the potential singular values  when the denominator $\left \langle S^x,S^y,S^z \right \rangle $ are zero. 

\begin{align}
{\cal {C}}_{\text {ent }} &\equiv  \frac{\text{Var}\left(S_{1}^{x}-S_{2}^{x}\right)+\text{Var}\left(S_{1}^{y}-S_{2}^{y}\right)+\text{Var}\left(S_{1}^{z}+S_{2}^{z}\right)}{4 N} \nonumber \\ 
&\textless 1 \hspace{1cm} \text{(for entangled states)}
\label{HT criteria}
\end{align}

Another criterion is the Duan-Giedke-Cirac-Zoller (DGCZ) criterion \cite{PhysRevLett.84.2722}, any violation of it implies the existence of entanglement

\begin{align}
{\cal {C}}_{\text {DGCZ }} &\equiv \frac{\text{Var}\left({S}_{1}^{y}-{S}_{2}^{y}\right)+\text{Var}\left({S}_{1}^{z}+{S}_{2}^{z}\right)}{2\left(\left|\left\langle S_{1}^{x}\right\rangle\right|+\left|\left\langle S_{2}^{x}\right\rangle\right|\right)} \nonumber \\
&\geqslant 1  \hspace{0.8cm} \text{(for separable states)}
\label{DGCZ criteria}
\end{align}

The third criterion we introduce here is the Wineland squeezing criterion \cite{PhysRevA.50.67}

\begin{align}
\xi^{2}&=\frac{2\left(\Delta S_{\perp}^{2}\right)_{\min }}{|\langle S\rangle|}  \nonumber \\
&\textless 1 \hspace{0.8cm} \text { (for squeezing states) }
\label{wineland}
\end{align}

In the given expression, $S_{\perp}$ (Appendix.\ref{wineland minimal}) signifies any component that is perpendicular to the mean total spin, while $\Delta S_{\perp}^{2}$ denotes the minimum fluctuation in the vertical direction.

EPR steering is developed by Reid and co-authors \cite{2018Demonstration,RevModPhys.81.1727}, and a violation of the inequality (\ref{EPR steering}) implies the existence of EPR steering from BEC 1 to BEC 2.  

\begin{align}
\mathcal{C}_{\text {steer }}^{1 \rightarrow 2} &\equiv \frac{\operatorname{Var}\left(S_{1}^{y}-S_{2}^{y}\right) \operatorname{Var}\left(S_{1}^{z}+S_{2}^{z}\right)}{\left\langle S_{1}^{x}\right\rangle^{2}} \nonumber \\
 &\geq 1 \quad \text { (un-steerable) }
\label{EPR steering}
\end{align}

\begin{figure}[t]
\centering
\includegraphics[width=\columnwidth]{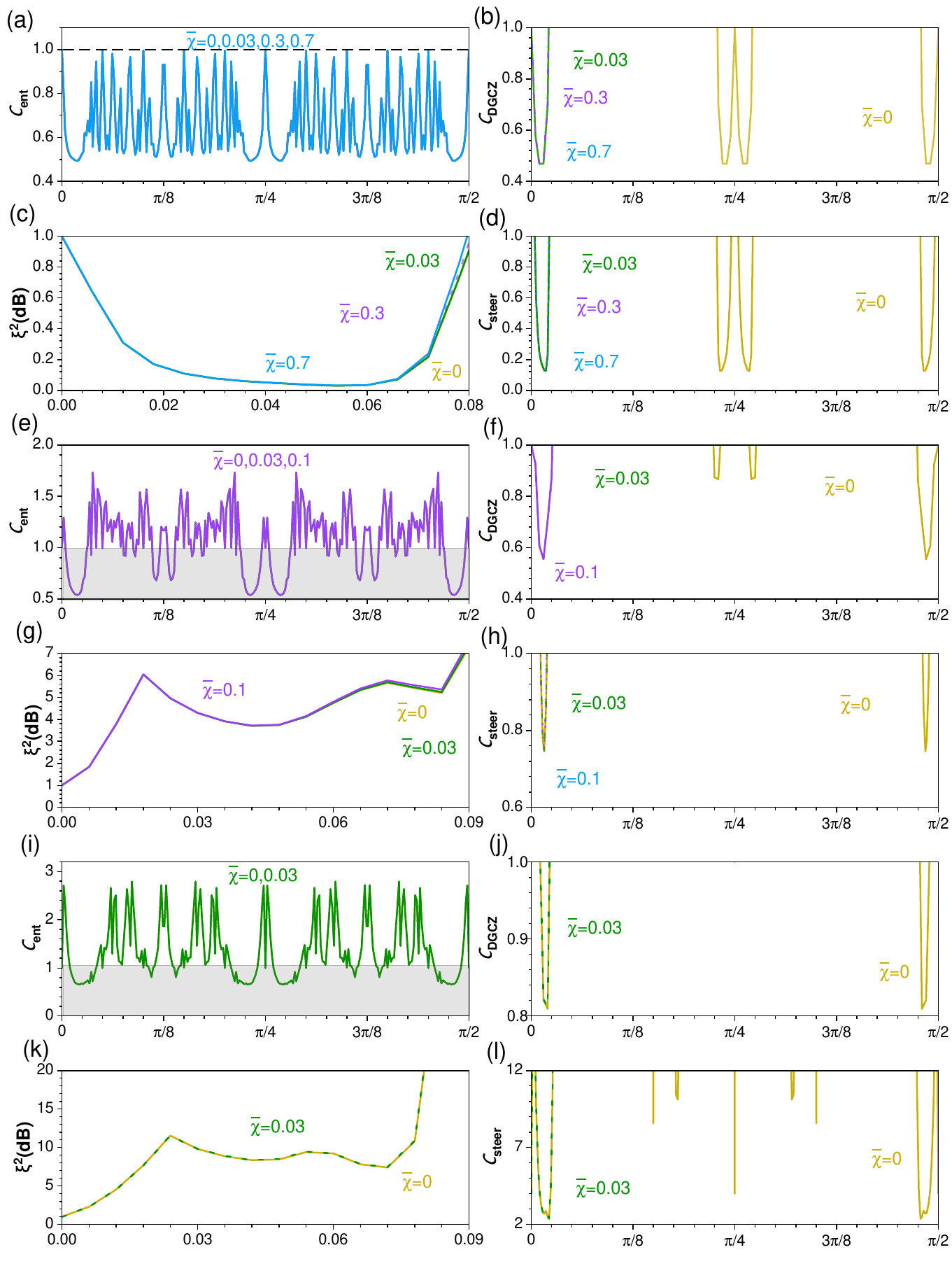}
\caption{Entanglement criteria for the state (\ref{eq:FinaleUnnormalized}) under photon loss channel (a,e,i) The Hofmann-Takeuchi criterion (\ref{HT criteria}) .vs interaction time $\tau$ with different amplitude attenuation coefficients (b,f,j) The DGCZ criterion (\ref{DGCZ criteria}) .vs interaction time $\tau$ with photon loss decoherence. (c,g,k) The Wineland squeezing criterion versus interaction time $\tau$ under photon loss channel. (d,h,l) The EPR steering criterion versus interaction time $\tau$ when photon loss decoherence is included. All the four plots in (a-d) used photon outcomes $n_d=n_c=50$, plots (e-h) adopted $n_d=60,n_c=40$ and $n_d=70,n_c=30$ for plots (i-l). The common parameters used are $N=20,\alpha=10$. The masked regions in (e,i) signify where correlation-based entanglement could be witnessed. The amplitude attenuation coefficients $\bar{\chi}$ are marked in plots. }
\label{entanglement_criteria}
\end{figure}

Fig.\ref{entanglement_criteria} shows the correlation-based criteria for the state (\ref{eq:FinaleUnnormalized}) under the photon loss channel. Fig.\ref{entanglement_criteria}(a-d) displays the Hofmann-Takeuchi criterion (\ref{HT criteria}), DGCZ criterion (\ref{DGCZ criteria}), Wineland squeezing (\ref{wineland}) and EPR steering criterion (\ref{EPR steering}) for photon measurement outcomes $n_d=n_c=50$, respectively. The black dashed line in Fig.\ref{entanglement_criteria}(a) indicates the entanglement boundary, which shows the region where entanglement can be detected. Comparing the regions detected by the HT criteria in the three cases, we observe that as $d$ mode photon numbers $n_d$ increases, a smaller time domain can be detected, and HT criteria is quite robust under photon loss decoherence. This is on account of $\text {Var} (S_1 ^ x- S_2 ^ x) + \text {Var} (S_1 ^ y - S_2 ^ y) \approx 2 N $ on the whole time regions, so the characteristics of the HT criterion is mainly determined by $\text {Var} (S_1^z+ S_2^z) $ which is invariant as shown in Fig.\ref{variance_all}(c,f,j). In general, the HT criterion is relatively stable for photon loss decoherence. The HT criterion can detect the largest time domain compared to other criteria.

The DGCZ criterion, defined by (\ref{DGCZ criteria}), is used as the second detection criterion to compare entanglement between different $d$ mode photon outcomes $n_d=50,60,70$. All three cases indicate that the DGCZ criterion can detect entanglement even with significant amplitude attenuation coefficients. For the $n_d=n_c=50$ case, the DGCZ criterion detects entanglement in the same time detection region as the EPR criterion. Turning to the $n_d=60,n_c=40$ case, the DGCZ criterion detects a wider time domain of entanglement than the EPR criterion. It originated from the EPR criterion is a higher level of entanglement detection criterion. In both $n_d=n_c=50$ and $n_d=60,n_c=40$ cases, the DGCZ criterion can detect entanglement over a broader time range when no photon loss is included. However, when photon loss decoherence is introduced, the DGCZ criterion only reports the existence of entanglement within a small time region. With more severe photon loss effects, the time intervals in which the DGCZ criterion is capable of detecting hardly change. It demonstrates that the DGCZ criterion seems to be highly suitable for detecting entangled states.

The EPR criterion is the third-best detection criterion of the four mentioned criteria. For $d$ mode photon outcomes with $n_d=50$ and $n_d=60$, it can detect entanglement even when there are strong photon losses. Regarding the $n_d=70$ case, it loses its effectiveness in capturing entanglement. Nevertheless, the region where entanglement could be witnessed in a short period does not significantly vary as the photon loss intensifies. Unfortunately, the Wineland squeezing criterion is only effective for the $n_d=50$ case, making it the worst detection criterion. As anticipated, the entanglement ranges captured by Wineland squeezing in Fig.\ref{entanglement_criteria}(c) decrease slightly as the photon loss increases. It loses its detection capability for $n_d=60,70$ cases, depicted in Fig.\ref{entanglement_criteria}(g,k).

%

%
%

\section{Conclusion}
\label{conclusion}

We have studied the effects of photon loss on the entangled state with the QND measurement scheme between BECs. By employing the IWOP and TESR techniques, we have been able to accurately determine the density matrix and investigate the evolution of states under various conditions, including coherent light intensity, interaction times, and amplitude attenuation coefficients. This approach offers significant advantages compared to traditional methods. We investigate the properties of entangled states under three different photon measurement outcomes and find that the entangled states become more exotic with unbalanced photon numbers in $c,d$ modes.

As the value of $d$ mode photon numbers $n_d$ increases, the Wigner distributions of the entangled state become more non-classical in $\tau = 1/N, 1/(2\sqrt{N})$. At $\tau \lesssim 1/N$ when HP approximation is effective, the Wigner distribution shows no differences under photon loss decoherence, which is consistent with that observed in Refs.\cite{PhysRevA.88.023609,Gao_2022}. There have been significant enhancements in entanglement values without decoherence at specific time points for cases $n_d=60,70$. While these cases are more susceptible to optical loss decoherence than the $n_d=50$ scenario, they still maintain a certain level of entanglement. The entangled states for all three cases possess correlations in the spin $S_x,S_y$ basis while anti-correlations in the spin $S_z$ basis.

The detection of entanglement through correlations has been widely examined using various criteria such as the HT, DGCZ, Wineland squeezing, and EPR steering criteria. Among these criteria, the HT criterion has the widest entanglement time domains and is very robust to different degrees of photon loss, making it the most effective tool in witnessing entanglement. The criteria are ranked in order of detection ability from best to worst: HT > DGCZ > EPR steering > Wineland squeezing. As photon loss decoherence increases over time, the detection range for the entanglement criterion remains relatively stable. This is an inspiring result from an experimental perspective.

\section*{Acknowledgments}

Numerical computations were performed at the Hefei Advanced Computing Center. This work is supported by the Primary Research $\&$ Development Plan of Jiangsu Province (Grant No. BE2016175). T.B. is supported by the
National Natural Science Foundation of China (62071301); the Science and Technology Commission of Shanghai Municipality (19XD1423000,
22ZR1444600); the China Science and Technology Exchange
Center (NGA-16-001); the China Foreign Experts Program
(G2021013002L). A.P. is supported by RSF (Grant No. 23-21-00507)

\subsection*{Conflict of Interests}
The authors declare that they have no known competing financial interests or personal relationships that could have appeared to influence the work reported in this paper.

\section*{Data availability}

Data that supports the findings of this study is available
from the corresponding author upon request.

\appendix

\section{Fock state matrix elements}
\label{Fock state elements}
The Fock states in bases $l,m \in \{ x,y \} $ are defined as 

\begin{align}
|k\rangle^{(x)} &=e^{-i S^{y} \pi / 4}|k\rangle^{(z)} \nonumber \\
|k\rangle^{(y)} &=e^{-i S^{z} \pi / 4} e^{-i S^{y} \pi / 4}|k\rangle^{(z)}
\label{fock state in various bases}
\end{align}

The matrix of the $S^y$ rotation is 
\begin{align}
& \langle k | e^{-i S^y \theta/2} | k' \rangle = \sqrt{ k'! (N-k')! k! (N-k)!} \nonumber \\
& \times 
\sum_n \frac{(-1)^n \cos^{ k- k' + N - 2n} (\theta/2) \sin^{2n + k' - k} (\theta/2) }{(k-n)!(N-k'-n)!n!(k'-k+n)!}, 
\label{syrotmatrixelement}
\end{align}
where $ | k \rangle $ is the eigenstates of $ S^z $. The matrix elements of $ S^x $ are accordingly given by
\begin{align}
& \langle k | e^{-i S^x \theta/2} | k' \rangle = 
i^{k'-k} \langle k | e^{-i S^y \theta/2} | k' \rangle,
\label{sxrotmatrixelement}
\end{align}
considering the fact $ S^x = e^{-i S^z \pi/4} S^y e^{i S^z \pi/4} $. 

\section{Minimum fluctuation}
\label{wineland minimal}

In the Bloch sphere, the mean spin direction, azimuth, and polarization angle $\theta,\phi$ are calculated by

\begin{align}
\boldsymbol{n}_{1} &=\sin \theta \cos \phi \left\langle S_{x}\right\rangle+\sin \theta \sin \phi \left\langle S_{y}\right\rangle+\cos \theta \left\langle S_{z}\right\rangle, \nonumber \\
\theta &=\arccos \frac{\left\langle S_{z}\right\rangle}{\sqrt{\left\langle S_{x}\right\rangle^{2}+\left\langle S_{y}\right\rangle^{2}+\left\langle S_{z}\right\rangle^{2}}}, \nonumber \\
\phi&=\arccos \frac{\left\langle S_{x}\right\rangle}{\sqrt{\left\langle S_{x}\right\rangle^{2}+\left\langle S_{y}\right\rangle^{2}}} .
\label{Bloch angles}
\end{align}

Since $\left\langle S_{y}\right\rangle=0$, it indicates the mean spin is in the $x-z$ plane, so the spin squeezing occurs in $(y,\boldsymbol{n}_{2})$ plane where

%
\begin{align}
\boldsymbol{n}_{2}=-\cos \theta \left\langle S_{x}\right\rangle +\sin \theta \left\langle S_{z}\right\rangle
\end{align}

The spin component $\left\langle S_\zeta\right\rangle$ of any angle $\zeta$ on $(y,\boldsymbol{n}_{2})$ plane is  defined as

\begin{align}
S_\zeta &=S_{y} \cos \zeta+S_{\boldsymbol{n}_{2}} \sin \zeta, \nonumber \\
& = S_{y} \cos \zeta -\cos \theta \sin \zeta S_x+\sin \theta \sin \zeta S_z
\end{align}

Then the minimum fluctuation in the vertical direction $\Delta S_{\perp}^{2}$ and the optimal angle $\zeta_{\text{opt}}$
are deduced in the following expressions

\begin{align}
\Delta S_{\perp}^{2}&=\left \langle{S_{\zeta^{\text{(opt)}}}^2} \right \rangle-{\left \langle{S_{\zeta^{\text{(opt)}}}} \right \rangle}^2, \nonumber \\
\zeta_{\text{opt}}&=\frac{1}{2}\left[\pi+\arctan \frac{\left\langle S_{y} S_{\boldsymbol{n}_{2}}+S_{\boldsymbol{n}_{2}} S_{y}\right\rangle}{\left\langle S_{y}^{2}-S_{\boldsymbol{n}_{2}}^{2}\right\rangle}\right]
\end{align}

\bibliographystyle{apsrev}
\bibliography{gaoshuai}

\end{document}